\documentclass[manuscript]{aastex}
\usepackage{amsmath}
\usepackage{url}

\def\deg{\ifmmode {^\circ}\else {$^\circ$}\fi}
\def\degree{\ifmmode {^\circ}\else {$^\circ$}\fi}
\def\mum{\ifmmode {\rm \,\mu {\rm m}}\else $\rm \,\mu {\rm m}$\fi}
\def\arcsec{\ifmmode ^{\prime \prime}\else $^{\prime \prime}$\fi}

\def\inch{\ifmmode ^{\prime \prime}\else $^{\prime \prime}$\fi}
\def\msunyr{\ifmmode {M_{\odot}~{\rm yr^{-1}}}\else $M_{\odot}~{\rm yr^{-1}}$\fi}
\def\msun{\ifmmode {M_{\odot}}\else $M_{\odot}$\fi}
\def\rsun{\ifmmode {R_{\odot}}\else $R_{\odot}$\fi}
\def\lsun{\ifmmode {L_{\odot}}\else $L_{\odot}$\fi}
\def\mstar{\ifmmode {M_{\star}}\else $M_{\star}$\fi}
\def\rstar{\ifmmode {R_{\star}}\else $R_{\star}$\fi}
\def\tstar{\ifmmode {T_{\star}}\else $T_{\star}$\fi}
\def\lstar{\ifmmode {L_{\star}}\else $L_{\star}$\fi}
\def\md{\ifmmode {M_d}\else $M_d$\fi}
\def\ld{\ifmmode {L_d}\else $L_d$\fi}
\def\ad{\ifmmode A_d\else $A_d$\fi}
\def\ldlstar{\ifmmode L_d / L_\star\else $L_d / L_{\star}$\fi}
\def\rearth{\ifmmode {\rm R_{\oplus}}\else $\rm R_{\oplus}$\fi}
\def\mearth{\ifmmode {\rm M_{\oplus}}\else $\rm M_{\oplus}$\fi}
\def\qdstar{\ifmmode Q_D^\star\else $Q_D^\star$\fi}
\def\kms{\ifmmode {\rm km~s^{-1}}\else $\rm km~s^{-1}$\fi}
\def\ms{\ifmmode {\rm m~s^{-1}}\else $\rm m~s^{-1}$\fi}
\def\mesc{\ifmmode m_{esc}\else $m_{esc}$\fi}
\def\rmin{\ifmmode r_{min}\else $r_{min}$\fi}
\def\rmax{\ifmmode r_{max}\else $r_{max}$\fi}
\def\mmin{\ifmmode m_{min}\else $m_{min}$\fi}
\def\mmax{\ifmmode m_{max}\else $m_{max}$\fi}
\def\rmind{\ifmmode r_{min,d}\else $r_{min,d}$\fi}
\def\rmaxd{\ifmmode r_{max,d}\else $r_{max,d}$\fi}
\def\mmaxd{\ifmmode m_{max,d}\else $m_{max,d}$\fi}
\def\vrad{\ifmmode v_{rad}\else $v_{rad}$\fi}
\def\qz{\ifmmode q_{0}\else $q_{0}$\fi}
\def\qi{\ifmmode q_{i}\else $q_{i}$\fi}
\def\ql{\ifmmode q_{l}\else $q_{l}$\fi}
\def\qs{\ifmmode q_{s}\else $q_{s}$\fi}
\def\rbrk{\ifmmode r_{brk}\else $r_{brk}$\fi}
\def\rdamp{\ifmmode r_{damp}\else $r_{damp}$\fi}
\def\rin{\ifmmode r_{in}\else $r_{in}$\fi}
\def\rout{\ifmmode r_{out}\else $r_{out}$\fi}
\def\tin{\ifmmode t_{in}\else $t_{in}$\fi}
\def\tout{\ifmmode t_{out}\else $t_{out}$\fi}
\def\ain{\ifmmode a_{in}\else $a_{in}$\fi}
\def\aout{\ifmmode a_{out}\else $a_{out}$\fi}
\def\r0{\ifmmode R_{0}\else $R_{0}$\fi}
\def\m0{\ifmmode m_{0}\else $m_{0}$\fi}
\def\M0{\ifmmode M_{0}\else $M_{0}$\fi}
\def\xm{\ifmmode x_{m}\else $x_{m}$\fi}
\def\sigz{\ifmmode \Sigma_0\else $\Sigma_0$\fi}
\def\gyr{\ifmmode {\rm g~yr^{-1}}\else ${\rm g~yr^{-1}}$\fi}
\def\cms{\ifmmode {\rm cm~s^{-1}}\else ${\rm cm~s^{-1}}$\fi}
\def\gcms{\ifmmode {\rm g~cm^{-2}}\else $\rm g~cm^{-2}$\fi}
\def\gcmss{\ifmmode {\rm g~cm^{-2}~s^{-1}}\else $\rm g~cm^{-2}~s^{-1}$\fi}
\def\gcmc{\ifmmode {\rm g~cm^{-3}}\else $\rm g~cm^{-3}$\fi}
\def\dcm2{\ifmmode {\rm dyn~cm^{-2}}\else $\rm dyn~cm^{-2}$\fi}
\def\ecsk{\ifmmode {\rm erg~cm^{-1}~s^{-1}~K^{-1}}\else $\rm erg~cm^{-1}~s^{-1}~K^{-1}$\fi}
\def\ec2s{\ifmmode {\rm erg~cm^{-2}~s^{-1}}\else $\rm erg~cm^{-2}~s^{-1}$\fi}
\def\cm2{\ifmmode {\rm cm^{-2}}\else $\rm cm^{-2}$\fi}
\def\atilin{\ifmmode {\tilde{a}_{in}}\else $\tilde{a}_{in}$\fi}
\def\atilout{\ifmmode {\tilde{a}_{out}}\else $\tilde{a}_{out}$\fi}
\def\atil{\ifmmode {\tilde{a}}\else $\tilde{a}$\fi}
\def\ttil{\ifmmode {\tilde{t}}\else $\tilde{t}$\fi}
\def\sqrttt{\ifmmode {\tilde{t}^{1/2}}\else $\tilde{t}^{1/2}$\fi}

\def\h2o{H$_2$O}
\def\sio2{SiO$_2$}

\shorttitle{Non-Adiabatic Thermal Profiles}
\shortauthors{Podolak, Helled, Schubert}

\begin{document}

\title{Effect of Non-Adiabatic Thermal Profiles on the Inferred Compositions of Uranus and Neptune}

\author{Morris Podolak$^1$, Ravit Helled$^2$ and Gerald Schubert$^3$}
\affil{$^1$Department of Geoscience,\\
Tel-Aviv University, Tel Aviv, Israel. \\
$^2$Center for Theoretical Physics and Cosmology, \\
University of Zurich, Zurich, Switzerland. \\
$^3$Department of Earth, Planetary and Space Sciences, \\
University of California, Los Angeles, USA. \\
}

\begin{abstract}
It has been a common assumption of interior models that the outer planets of our solar system are convective, and that the internal temperature distributions are therefore adiabatic. This assumption is also often applied to exoplanets. However, if a large portion of the thermal flux can be transferred by conduction, or if convection is inhibited, the thermal profile could be substantially different and would therefore affect the inferred  planetary composition.  Here we investigate how the assumption of non-adiabatic temperature profiles in Uranus and Neptune affects their internal structures and compositions. We use a set of plausible temperature profiles together with density profiles that match the measured gravitational fields to derive the planets' compositions. We find that the inferred compositions of both Uranus and Neptune are quite sensitive to the assumed thermal profile in the outer layers, but relatively insensitive to the thermal profile in the central, high pressure region. The overall value of the heavy element mass fraction, $Z$, for these planets is between 0.8 and 0.9.  Finally, we suggest that large parts of Uranus' interior might be conductive, a conclusion that is consistent with Uranus dynamo models and a hot central inner region.  
\end{abstract}


\section{Introduction}
One of the most fundamental and interesting pieces of information that we can derive for a planet is its overall metallicity, $Z$.  Measurements of a planet's mass, radius, and gravitational moments are not sufficient to uniquely fix its internal density distribution \citep{marley95, podpod00}.  Even if such a density distribution is obtained, interpreting that density in terms of composition requires a knowledge of the planet's thermal profile.  For planets in our solar system, where the mass, radius, and gravitational field are well characterized, the thermal profile remains ambiguous.  For Jupiter it was long thought that this problem is not very acute and simple arguments could be made to strongly constrain the thermal profile and general composition.  Jupiter's low density requires that a large fraction of its mass be composed of hydrogen and helium, while its high thermal flux requires that much of its volume be convective \citep{hubbard68}, and this implies that much of the interior follows an adiabatic temperature profile.  However, more recent work questions this view.  The latest measurements of Jupiter's gravity field by the Juno spacecraft have been interpreted to imply that there are discontinuities in composition and entropy in Jupiter's interior \citep{wahl2017,debras2019}.  Such discontinuities had been suggested earlier on the basis of evolutionary considerations \citep[see, e.g.][]{nettelmann15, mankovich2016, vazan2018}, so that a simple convective interior is not a good approximation for Jupiter. 

Similar arguments can be made for Saturn, although they are somewhat less convincing, and the case for a fully-convecting Saturn is more problematic.  As is well-known, the dipole of Saturn's magnetic field is nearly aligned with its rotation axis.  This can be explained by assuming that there is a stably stratified conducting layer overlying the dynamo region \citep{stevenson82b}.  A recent analysis of Cassini data puts the thickness of this layer at $\geq 4000$\,km \citep{cao2011}.  In addition, \cite{leconte13} have argued that diffusive regions might be needed to explain Saturn's current luminosity, which is higher than expected from simple cooling models.  Such diffusive regions might occur in both Jupiter and Saturn where convection could be inhibited due to the separation of helium from hydrogen and its subsequent rainout \citep{stevsal77}. 

Convection can also be inhibited in the gas giants by compositional gradients that can result from the formation process itself.  Such compositional gradients can form as accreted planetesimals dissolve in different regions of the protoplanet \citep[e.g.][]{Lozovsky17}, or if the core erodes due to miscibility effects \citep[e.g.][]{wilson12}.  Thus the assumption of a fully convective Jupiter and Saturn is less obvious than one might expect.  Uranus and Neptune are not simply smaller versions of Jupiter and Saturn.  Their internal compositions are notably different, and the arguments for their interiors being convective are correspondingly weaker.  This is particularly true for Uranus whose internal heat source is close to zero.  More details on non-adiabaticity of the outer planets can be found in \citet{HG2017}. 

One of the reasons that the assumption of an adiabatic interior is popular, is that invoking it precludes the need to model the thermal flux.  The temperature profile can be computed simply by computing the adiabatic heating due to compression of the material.  As a result, adiabatic interior models do not require a thermal flux calculation.  Even evolutionary models \citep{hubmac80, fortnet10} sidestep the problem of calculating the thermal flux by tying the effective temperature at which the planet radiates to the interior adiabat.  In this way the luminosity of the planet is linked to its total heat content and a cooling curve is calculated without actually computing the heat transport within the planet.  If the planet is not convecting throughout the interior, however, the adiabatic assumption fails and the derived thermal structure is inappropriate.  The correspondence between the pressure-density relation derived from fitting the gravitational field and the pressure-density relation computed for a particular composition is established via the associated temperature profile.  Therefore it is important to determine how much this thermal profile can vary from the adiabat that is commonly assumed, and how much this variation is likely to affect the inferred value of $Z$.  

Both Uranus and Neptune show evidence that their interiors might not be adiabatic.  With adiabatic models Uranus requires billions of years more than the age of the solar system to reach its present state \citep{fortnet10}.  Models of their evolution show that although Neptune can cool to its present state within the lifetime of the solar system, it would have to have started with much lower internal temperatures than are predicted by current formation scenarios.  Uranus is a particularly interesting case.  Its low thermal flux \citep{pearl90} suggests that either it formed with a cool initial thermal profile, and underwent further cooling to come to thermal equilibrium with the solar insolation, or that there is a mechanism, such as a thermal boundary layer \citep[see, e.g.][]{Nett2016}, which is preventing the heat from escaping efficiently.  \cite{Nett2016} found that the existence of such a boundary layer can explain Uranus' low luminosity, leading to a hotter interior which could consist of more refractory materials. Additionally, there might be some mechanism, such as layered convection and/or conductive/radiative regions \citep[e.g.][]{leconte13,Vazan2016} that is preventing the heat from escaping efficiently.  In all of these cases the thermal profile may differ substantially from an adiabat.

For both Uranus and Neptune hydrogen and helium are relatively minor components, and most of the mass consists of heavier materials.  It is not clear exactly how to choose and arrange these materials in the interior.  Traditionally, models have been based on the assumption that these planets consist of a core of rocky material surrounded by an ice shell which is overlain by a hydrogen-helium atmosphere.  Such models succeed in fitting the observed gravity field with a composition that is consistent with the formation scenarios of these planets \citep{podolak84, podolak91, podolak95, nettel12b}.  Other models have suggested that the transition from a hydrogen-helium rich atmosphere to an ice-rich shell is more gradual \citep{marley95, helledetal10}.  However, as \cite{podpod00} have shown, many other density distributions are consistent with the gravity field.  Some of these are clearly unphysical and can be discarded, but others may represent plausible alternative compositions.  Overall, standard structure models of Uranus and Neptune suggest bulk metallicities of 0.75--0.92 and 0.76--0.9, and minimum H-He masses of 2 and 3 M$_\oplus$, respectively \citep[see][and references therein]{HG2017}.

The interpretation of the magnetic fields of Uranus and Neptune is also dependent on the assumed thermal structure.  In order to fit the measurements an off-center dipole that is strongly inclined to the rotation axis is required.  Can the proposed composition supply the required conductivity to generate such an off-center dipole?  Improved knowledge of the conductivity of relevant materials  \citep[e.g.][]{redmer11, nellis2015, Kraus2017} as well as better models of magnetic field generation \citep[e.g.][]{stanblox04,stanblox06} allow us to assess the sources of the magnetic fields of those planets provided we can put useful limits on their compositions and temperatures.  As a result it is important to understand how well we know the structure and composition of Uranus and Neptune, and, in particular, how sensitive the model results are to the assumed thermal profile.  Finally, understanding the structure of Uranus and Neptune has become a particularly timely subject in view of the large number of exoplanets that have been discovered with radii of 2-6 R$_{\oplus}$ \citep[see, e.g.][]{petigura2013, Weiss2014, Zeng2016, Fulton2017}.  Attempts at classifying their compositions in terms of a mass-radius diagram rely, in no small part, on inferences drawn from the structure of the ice giants in our own solar system \citep{lopez2014}.  

In what follows, in section 2 we summarize the physical background for the models.  In section 3 we present the formalism we use to generate thermal profiles for Uranus and Neptune. In section 4 we apply this formalism and derive values of $Z$ for different Neptune profiles. The interesting case of Uranus, with its low thermal flux is presented in section 5, and our summary and conclusions are given in section 6.

\section{Physical Background}

\subsection{Thermal Profile}
Given a density profile that fits the observed gravitational field of a planet, the corresponding pressure profile can be computed from the hydrostatic equilibrium equation. Together with a temperature profile, a self-consistent (but not unique!) composition can be found.  Without the adiabatic assumption, however, computing a temperature profile is difficult.  

\subsection{Thermal Flux Models}
First, it is necessary to know the magnitude of the thermal flux that must be transported.  The problem is that the thermal flux can be directly measured only at the planetary surface.  The interior flux, which is required in order to compute a thermal gradient in the non-convecting case, can only be inferred indirectly.  Below we consider two simple models for the flux.  The rate of energy entering a shell of radius $r$ and thickness $dr$ is $$\frac{dE_{in}}{dt}=4\pi r^2F(r)$$ where $F$ is the thermal flux. The rate of energy leaving the shell is the rate of energy entering the shell plus the rate of energy produced inside the shell $$\frac{dE_{out}}{dt}=4\pi r^2F(r)+4\pi r^2 drc_v\rho \dot{T}=4\pi (r+dr)^2 F(r+dr)\approx 4\pi r^2\left(1+2\frac{dr}{r}\right)(F+dF)$$
To first order this gives the differential equation,
\begin{equation}\label{difeq}
\frac{dF}{dr}+2\frac{F}{r}-c_v\rho\dot{T}=0.
\end{equation}
If the internal density $\rho$, heat capacity $c_v$, and cooling rate $\dot{T}$, are constant and independent of radius, the solution is
\begin{equation}\label{sol1}
F(r)=\frac{c_v\rho\dot{T}}{3}r=F_0\frac{r}{R},
\end{equation}
where $R$ is the planetary radius and $F_0$ is the surface flux. Thus, the first model we consider is where the flux varies linearly with radius inside the planet.

The second case we consider is a planet whose central region, up to a radius $r_c$, is still cooling with constant $\dot{T}$, but for $r>r_c$ the temperature has reached some equilibrium value.  Thus for $r>r_c$ we have $\dot{T}=0$.  In this case the solution to Eq.\,\ref{difeq} is
$$F=\frac{c_v\rho\dot{T}}{3}r\ \ \  0\leq r\leq r_c$$ 
\begin{equation}\label{sol2}
=\frac{c_v\rho\dot{T}}{3}\frac{r_c^3}{r^2}\ \ \ \ r_c<r\leq R
\end{equation}
At the surface $$F(R)=F_0=\frac{c_v\rho\dot{T}}{3}\frac{r_c^3}{R^2}$$ so 
$$F(r)=F_0\frac{R^2}{r_c^3}r\ \ \ \ 0\leq r\leq r_c$$
\begin{equation}\label{solf0}
=F_0\left(\frac{R^2}{r^2}\right)\ \ \ \ \ \ r_c<r\leq R. 
\end{equation}
Below, in addition to models with $r_c=R$ we also consider $r_c=0.1R$ and $r_c=0.01R$.

\subsection{Heat Transport Mechanism}
It is also necessary to determine the means of energy transport as well as the parameters (opacity, thermal conductivity, etc.) that characterize that transport.  A convenient way of describing the process has been developed by \cite[][hereafter, LC12]{leconte12}. In this study we use the formalism developed by LC12 to generate temperature profiles for Uranus and Neptune.  We then use these profiles together with density distributions that fit the observed gravitational fields of the planets to explore the range of plausible compositions.  

It should be noted that the specific process of double-diffusive convection modeled by LC12 has not been demonstrated to exist in Uranus and Neptune.  The details of the transport depend on a number of parameters such as the Prandtl number, and the diffusivity ratio \citep[see, e.g.][]{mirouh12,wood13} which are not sufficiently well known.  In the following, we use the LC12 formalism as a plausible and convenient way to generate temperature profiles.  These are meant to serve as a guideline for the importance of thermal effects.  We make no other assertions regarding the details of the heat transport mechanism.

\section{The Thermal Profile Calculation}
Compositional gradients can inhibit convection, and in such a case the temperature gradient can be super-adiabatic.  LC12 have developed a formalism for computing the temperature gradient when compositional gradients exist.  Their model assumes that although convection does occur, it is restricted to relatively narrow regions that are separated by thin diffusive layers.  The convective regions follow a nearly adiabatic profile over which the temperature change is small, while the diffusive regions have larger temperature gradients across them.  The LC12 prescription allows for a convenient averaging over these two types of regions.  The rate of heat transport depends on a number of parameters, including the number of layers assumed.  As noted above, we do not make any claims as to whether this mechanism is, indeed, present.  We simply adopt the LC12 method as a convenient way for parameterizing the heat transport through a region that is not fully convective.

The average temperature gradient can be written as \citep{leconte12}:
\begin{equation}\label{tgrad}
\frac{d\langle T\rangle}{dP}=\frac{T}{P}\langle\nabla_T\rangle=\frac{T}{P}\left\lbrace\nabla_{ad}+(\nabla_d-\nabla_{ad})\left[\left(\frac{\Phi}{C_L}\right)^{-1/4(1+a)}+\left(\Phi C_L^{1/a}\right)^{-a/(a+1)}\right]\right\rbrace
\end{equation}
where $\Phi$ is given by
\begin{equation}\label{phi2}
\Phi=\frac{\alpha_TgH_P^3}{\kappa_T^2}\left(\frac{\ell}{H_P}\right)^4\left[\frac{H_P}{\kappa_T\rho c_PT}F_{tot}-\frac{\alpha_T P}{\rho c_PT}\right],
\end{equation}
with $g$ being the acceleration of gravity and $\ell$ is the thickness of a convective layer.  $C_L$ and $a$ are constants which, following LC12, are set to be 1 and 0.3, respectively.  

The planet is divided  into a number of layers $N$ so that if $R$ is the total radius of the planet, $\ell=R/N$.  Further details can be found in LC12. The different $\nabla$'s refer to gradients due to different transport processes, where 
\begin{equation}\label{nabt}
\nabla_T\equiv\frac{d\ln T}{d\ln P}
\end{equation}
with $T$ and $P$ being the temperature and pressure, respectively.  $\nabla_{ad}$ is given by 
\begin{equation}\label{nabad}
\nabla_{ad}\equiv\left(\frac{\partial\ln T}{\partial\ln P}\right)_S=\frac{\alpha_T P}{\rho c_PT}
\end{equation} 
and $\nabla_d$ is an effective diffusive gradient given by 
\begin{equation}\label{nabd}
\nabla_d=F_{tot}\frac{H_P}{\kappa_T\rho c_PT}
\end{equation}
where $\alpha_T$ is the thermal expansivity, $\rho$ is the density, $c_P$ is the heat capacity at constant pressure, $H_P$ is the pressure scale height, $\kappa_T$ is the thermal diffusivity, and $F_{tot}$ is the total flux being transported. 

In addition, there are circumstances where the thermal gradient will be sub-adiabatic.  In this case, the system can maintain the necessary heat flux without recourse to convection and the LC12 picture does not apply.  This can happen in several settings, for example:
\begin{itemize}
\item{An atmosphere with a low opacity.  In this case the heat flux can be adequately transported by radiation.  We can approximate this within the LC12 formalism by an effective thermal diffusivity and still talk about an effective $\nabla_d$, thus retaining the LC12 formalism, if not their results.}
\item{A deep interior that has a high thermal conductivity.  In this case again the heat is carried by conduction and the thermal diffusivity is high.}
\item{A very low flux.  In this case even a low diffusivity might be enough to transport the energy at the required rate without recourse to convection.}
\end{itemize}
The common ingredient in the above cases is that $\nabla_d<\nabla_{ad}$.  This causes $\Phi$ in Eq.\,\ref{phi2} to become negative, and the LC12 formalism breaks down. This is because in these cases the system is Schwarzschild stable and little, if any, of the heat is carried by convection.  In addition to the obvious factors that influence the efficiency of diffusive heat transport, such as $\kappa_T$ and $F_{tot}$, the isothermal expansivity also comes into play.  At high pressures, where $\alpha_T$ is low, the small gain in volume due to a temperature increase does not add much buoyancy to the material and convection is less efficient.  The temperature profile that is derived for a particular model therefore depends on the choices of the different parameters, as well as the assumed density profile.  

\subsection{Thermal Diffusivity}
The thermal diffusivity $\kappa_T$ is a complex function of the composition, pressure, and temperature.  While a constant value for $\kappa_T$ may be adequate for Jupiter and Saturn where a large fraction of the mass is a solar mix of hydrogen and helium as assumed by \citep[LC12,][]{leconte13}, this is less suitable for Uranus and Neptune.  In order to explore the effect of different compositions, we have considered three sets of parameters corresponding to rocky material (\sio2 for $r<R_{Z1}$), icy material (\h2o for $R_{Z1}\leq r\leq R_{Z2}$) and a solar mix of hydrogen and helium (for $r>R_{Z2})$.  For each of these materials the thermal conductivity is taken directly from the tables of \cite{potekhin99}. These tables were originally developed for neutron stars but they extend to the range relevant to planetary interiors.  The tables go down only to a temperature of 1000\,K, and below this temperature we set the conductivity to be $10^4$\,erg\,cm$^{-1}$\,s$^{-1}$\,K$^{-1}$, corresponding to $\kappa_T=10^4/\rho c_P$ cm$^2$ s$^{-1}$.  Further details can be found in \citep{potekhin2015}. Recently {\it ab initio} calculations have been performed, based on DFT, for the electronic \citep{french2017} and ionic \citep{french2019} conductivity of water.  For pressures and temperatures relevant to the deeper regions of the ice shell in Uranus and Neptune, we find that the DFT calculations give conductivities that are between one and two orders of magnitude lower than the values we used. As a result, the \cite{potekhin99} values should be viewed with caution, and probably represent an upper bound. Since the shells in Uranus and Neptune are unlikely to consist of pure water, it is hard to estimate the conductivity in this region. We note that additional DFT calculations of icy mixtures, such as water, ammonia, methane, etc., are desirable, and can be used to further constrain the thermal profiles of the ice giants.  Values of the thermal conductivity for different Uranus and Neptune models are given in Table\,\ref{tab:condt}.

\subsection{Isothermal Expansivity and Heat Capacity}
Similarly, for the isothermal expansivity, $\alpha_T$, and the heat capacity, $c_P$, we refined the assumption of LC12 of constant values, and computed them directly from the equation of state.  For $r>R_{Z2}$ we used the equation of state of hydrogen taken from the tables of \cite{saumchab95}, and computed $\alpha_T$ as a function of pressure along an adiabatic profile. For $R_{Z1}\leq r \leq R_{Z2}$ we calculated $\alpha_T$ from the equation of state for \h2o using the quotidian equation of state of \cite{vazan13}.  For $r<R_{Z1}$ we used the quotidian equation of state for \sio2.

\subsection{Effective Diffusion Coefficient in a Radiative Zone}
In a region where the opacity is sufficiently low, such as the upper atmosphere of the planet, the thermal flux is carried by radiation.  In this case it is the opacity rather than the conductivity that is important.  Here too the exact value depends on pressure, temperature, and composition.  In fact, both the composition and its phase must be considered.  If any of the material is in the form of grains, information on the grain size distribution is also required.  A detailed opacity calculation is beyond the scope of this study, and here we use a simplification. If the luminosity in the region is $L$ and the opacity is $\chi$, then the temperature gradient is given by \citep[see, e.g.][]{clayton68}, 
$$\frac{dT}{dr}=-\frac{3\chi \rho L}{64\pi r^2\sigma T^3}$$ where $\sigma$ is the Stefan-Boltzmann constant.  The luminosity is related to the flux by $L=4\pi r^2 F$ so the temperature gradient can be written as 
$$\frac{dT}{dr}=-\frac{3\chi \rho F}{16 \sigma T^3}$$ 

If energy is transported by conduction, then substituting $$\frac{dT}{dr}=\frac{dT}{dP}\frac{dP}{dr}$$ in Eq.\,\ref{nabd} gives
$$\frac{dT}{dr}=-\frac{F}{\rho c_P \kappa_T}$$ and equating these two expressions gives an effective $\kappa_T$ of
\begin{equation}\label{opac} 
\kappa_T=\frac{16 \sigma T^3}{3\chi c_P\rho^2}.
\end{equation}
For $c_P=1.5\times 10^8$, $\rho=4\times 10^{-4}$, $T=70$ all in cgs units then $\kappa_T=4/\chi$.  The atmospheric opacity is not simple to calculate, in particular because of the expected condensate clouds.  In order to estimate the influence of this effective $\kappa_T$ we take $\chi\sim 1$, which is the opacity expected for a solar mixture of gas and grains at $T\sim 100$\,K \citep{pollack85}.  This gives an effective $\kappa_T=4$, which is about 400 times larger than the value of 0.01 assumed by LC12.  In these models we assume this effective $\kappa_T$ in the upper atmosphere up to a pressure, $P_{rad}$.  Above this pressure we use the LC12 formalism described above.  In this work we assume $P_{rad}=10$\,bar.   

\subsection{Density Profile}
\begin{figure}[ht]
\centerline{\includegraphics[angle=0, width=17cm]{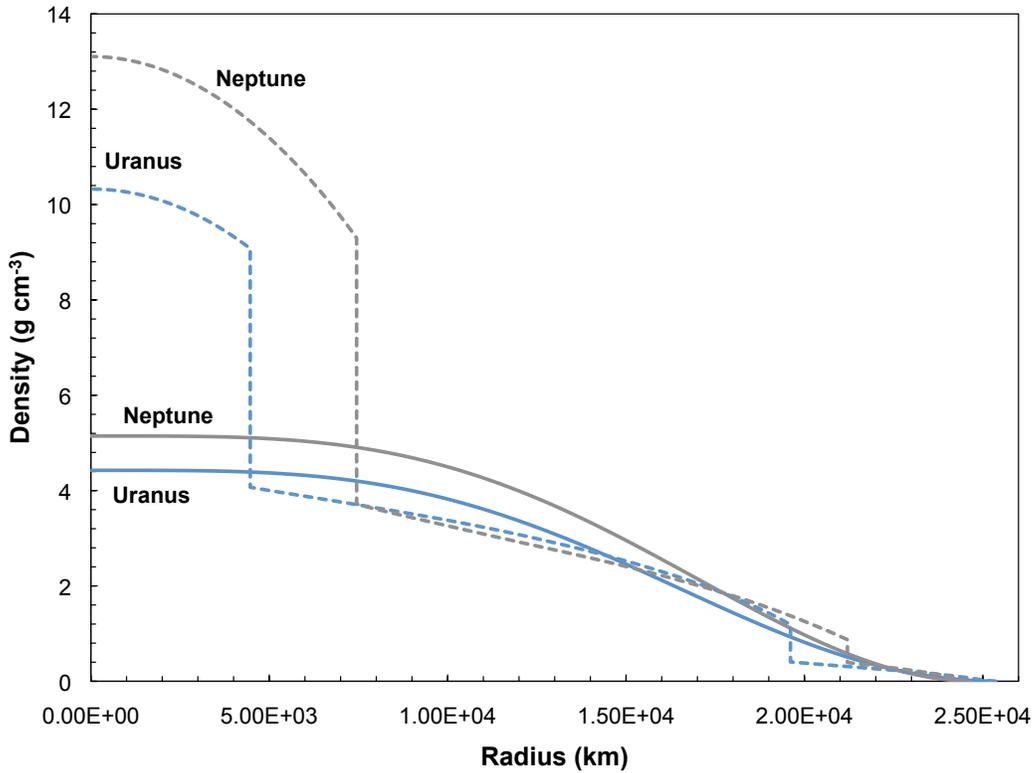}}
\caption{Density as a function of radius for Neptune (grey) and Uranus (blue). The solid curves are the density profiles presented in \cite{helledetal10}. The dashed curves are for the three-layer models of \cite{nettel12b}.}\label{denprof}
\end{figure}

For the density profiles of Uranus and Neptune we use the three-layer models U1 and N1 of \citep{nettel12b}. For comparison, we also investigate the polynomial density profiles presented in \cite{helledetal10}.  These density profiles are a sixth-order polynomials where the coefficients have been chosen to fit Uranus' Neptune's gravity field.  They contain no other assumptions regarding structure or composition.  The density profiles for these cases are shown in Fig.\,\ref{denprof}.  The higher order moments of the gravity field, $J_2$ and $J_4$ are strongly influenced by the density distribution in the outer parts of the planet, and are insensitive to the core.  As a result, very different central densities are possible.  However, since the density distribution must also reproduce the correct mass, high densities in the core must be offset by lower densities nearer the surface.  As discussed above, many other density profiles are conceivable, but the three-layer profile has the advantage of being based on self-consistent modeling using realistic composition considerations, while the polynomial profiles allow us to investigate a simple, but significant, deviation from the standard picture: a planet without a density discontinuity near the center.  

\section{Neptune}
We now proceed as follows:  Using a density profile which fits the measured gravitational parameters we use the LC12 formalism to generate a series of plausible thermal profiles.  The derived temperature is then used together with the density to infer a composition at different points in the planet.  If the composition is self-consistent, that is, it leads to a physically plausible structure, then the thermal profile is a reasonable one for this density profile; otherwise it must be discarded.  For the models of \cite{nettel12b} an adiabatic profile must give a self-consistent composition since this is how the models were originally computed, however for the models of \cite{helledetal10} we have no {\it a priori} expectations for what that profile might be.  What we wish to assess is whether non-adiabatic profiles can also be associated with a given density distribution, and how much such non-adiabatic profiles affect the inferred composition.

\begin{figure}[ht]
\centerline{\includegraphics[angle=0, width=17cm]{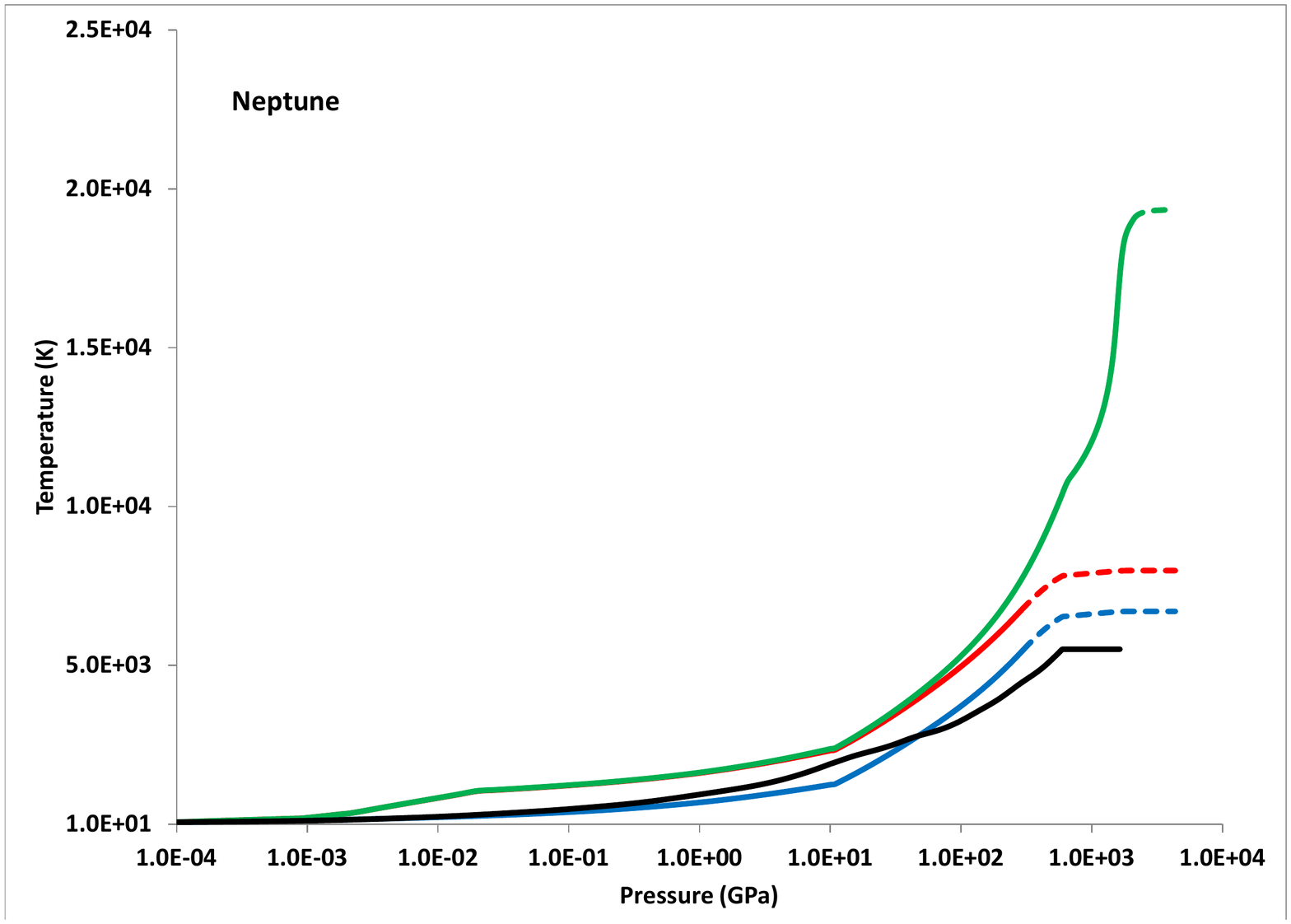}}
\caption{Thermal profiles as a function of internal pressure for Neptune models.  Shown are a 1 layer (blue) and a $10^6$ layer (red) model with $r_c=R$ and a $10^6$ layer model with $r_c=0.1R$ (green).  The temperature profile found by \cite{nettel12b} is shown in black. The solid parts of the curves are convective regions and the dashed parts are conductive.  See text for details. }\label{tnept}
\end{figure}

For the first set of thermal profiles for Neptune, we used the density profile for the planet from model N1 of \cite{nettel12b}.  We took $R_{Z1}=7.5\times 10^8$\,cm, corresponding to the radius of the core in those models, and $R_{Z2}$  = 2.1$\times10^9$\, cm corresponding to the outer radius of the ice shell.  We considered two profiles which are intended to sample the expected range of possible temperatures.  The first is characterized by $\ell=R$ in Eq.\,\ref{phi2} (1 layer model).  This corresponds roughly to a fully adiabatic profile, except that the adiabat is computed from the values we have chosen for the free parameters of the model rather than directly from the EOS.  The high temperature profile is characterized by choosing a small value for $\ell$.  There is no simple way to determine the smallest value that $\ell$ can take.  LC12 have argued that for Jupiter $\ell>4\times 10^{-6}R_{Jupiter}$, while for Saturn $\ell>3\times 10^{-5}R_{Saturn}$.  We therefore take $\ell=10^{-6}R$ ($10^6$ layer model) to compute the high temperature case.  The results are shown in Fig.\,\ref{tnept}.  The adiabat of \cite{nettel12b}, computed directly from the EOS, is shown in black.  The 1 layer model (blue) and the $10^6$ layer model (red) are shown for comparison.  The solid parts of the curve indicate convective regions, while the dashed parts show where conduction dominates.  These profiles essentially bracket that of \cite{nettel12b} at lower pressures. Above around 100\,GPa the temperature rises faster than the \cite{nettel12b} model.  For all cases the temperature profile was computed assuming a linear flux (Eq.\,\ref{sol1}) with $F_0=433$ \ec2s.  Both of the profiles we generated started with a 1-bar temperature equal to that of the \cite{nettel12b} model, 72\,K. Both were found to be convective up to a pressure of ~300\,GPa and became diffusive at higher pressures.  This is because of the higher conductivity in the high-pressure region of the ice layer.

\begin{figure}[ht]
\centerline{\includegraphics[angle=0, width=17cm]{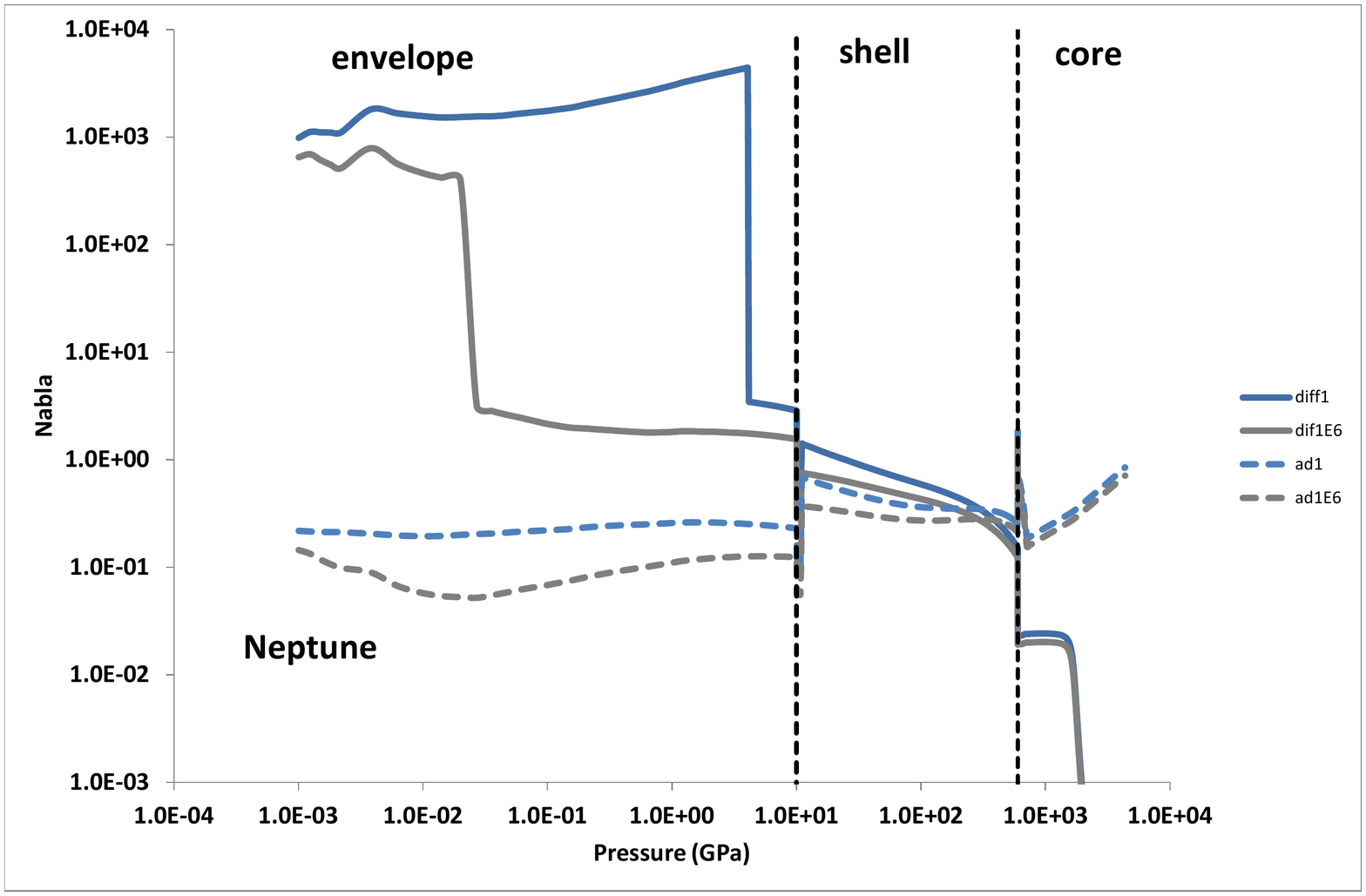}}
\caption{$\nabla_d$ (blue) and $\nabla_{ad}$ (grey) for a 1-layer (solid) and $10^6$-layer (dashed) model of Neptune assuming the density structure of \cite{nettel12b}. The black vertical dashed lines mark the envelope-shell and shell-core boundaries. See text for details. }\label{nabnep1}
\end{figure}

$\nabla_d$ and $\nabla_{ad}$ for both the 1-layer and the $10^6$-layer models are shown in Fig.\,\ref{nabnep1}.  The actual value used in computing $\nabla_T$ is the lower of the two.  As can be seen from the figure, for the envelope and much of the intermediate shell we find that $\nabla_T=\nabla_{ad}$. However, near the bottom of the shell and throughout the core $\nabla_T=\nabla_d$ and conduction carries heat more efficiently than convection.  This is because in this region the conductivity is sufficiently high, and at the same time the flux near the center is relatively low (Eq.\,\ref{sol1}). As a result, convection is not necessary to transport the flux. Finally, it is interesting to note that, aside from the ice shell, the diffusive and adiabatic gradients differ by more than an order of magnitude.  In addition, the conducting region is nearly the same for both models.

The models described above were computed assuming the flux is proportional to the radius of the planet (Eq.\,\ref{sol1}).   If, however, there is a process that prevents the heat from escaping from the inner region, it is possible that the outer region of the planet ($r>r_c$) would have cooled completely and the total energy passing through this region is constant.  In that case the flux is given by Eq.\,\ref{solf0}.  To explore such a scenario, we ran the same models taking the critical radius in Eq.\,\ref{solf0} to be $r_c=0.1R$.  We got a very similar structure for the envelope and the outer part of the shell (green curve in Fig.\,\ref{tnept}).  The temperatures differed by less than 10\% for pressures below about 500 GPa.  Above these pressures, however, the temperature profiles rose more quickly due to the higher thermal flux that has to be transported.  Central temperatures in the $10^6$-layer model reached $1.9\times 10^4$\,K as compared to $7.99\times 10^3$\,K for the constant flux case.  In addition, because of the higher flux near the center, the transport is not diffusive until pressures of 2\,TPa were reached, well within the core.

\subsection{Inferred composition} 
Once we have a thermal profile associated with the density and pressure profiles, we can use an EOS to calculate the density of any material under those conditions.  In particular, we consider hydrogen, helium, and a high-Z material, which we generally took to be \h2o.  We also ran some comparison cases using \sio2 or Fe as the high-Z material, but unless otherwise stated, the high-Z material is \h2o.  The mass fractions of hydrogen, helium, and the high-Z material are denoted by $X$, $Y$, and $Z$, respectively.  The ratio of hydrogen to helium is taken to be fixed at the solar ratio $Y=0.342X$ \citep{Lodders2010}, and we assume that the volume of a mixture of materials is equal to the sum of the volumes of the individual components.  Thus, the density $\rho$ of the mixture is given by $$\frac{1}{\rho}=\frac{X}{\rho_{H}}+\frac{Y}{\rho_{He}}+ \frac{Z}{\rho_Z}$$ The mass fractions have to sum to unity, so $Z=1-1.342X$. The density of the mixture is then given by, 
\begin{equation}\label{advol}
\frac{1}{\rho}=\frac{.745(1-Z)}{\rho_{H}}+\frac{.255(1-Z)}{\rho_{He}}+ \frac{Z}{\rho_Z}.
\end{equation}
Since the densities of the individual components are known, we can determine the value of $Z$ that matches the required density $\rho$ at each radius.  

It is important to keep in mind that if $Z$ is small, the most important contribution to Eq.\,\ref{advol} comes from the first term on the RHS both because of its large numerator and its small denominator.  As a result, small differences in the EOS for hydrogen or in the assumed H/He ratio can lead to large differences in the deduced value of $Z$  \citep[e.g.,][and references therein]{Miguel2016}.  Therefore small differences between the EOS used here and that of \cite{nettel12b} can cause noticeable differences in the estimated value of $Z$.  As a result, in such cases one must be careful not to over-interpret the inferred composition. However, the {\it trends} and sensitivities we find, are robust.  

To gauge sensitivity to the assumed EOS we ran cases where we used the SCVH EOS for hydrogen and helium \citep{saumchab95} and the quotidian EOS for the high-Z material \citep{vazan13}.  We also ran cases using the corresponding EOS tables taken from the SESAME data base \citep{sesame92} for comparison.  The differences in the computed value of $Z$ do not exceed 10\% and are usually much lower (only a few percent).  The results are summarized in Table\,\ref{tab:nadnep}. The models are shown only for the case of $r_c=R$ since the much higher temperatures generated for the case of $r_c=0.1R$ are limited to the highest pressures where $Z$ is much less sensitive to temperature. We show the results for the SESAME EOS, since the SCVH EOS tables do not include the lowest pressures and temperatures found in Uranus and Neptune.  Finally, we take the high-Z material to be \h2o throughout except for the core.  There we find that even pure \h2o does not reach a high enough density, so we use Fe for the high-Z material in this region.    

Looking at Table\,\ref{tab:nadnep}, we can see that although the inferred $Z$ in the envelope is quite constant for the \cite{nettel12b} profile, as expected, it oscillates for the other two profiles and implies that these are not consistent with the density profile.  In the envelope the $N=1$ (cold) profile is simply an approximation to the adiabatic profile within the boundaries of the LC formalism and begins to deviate from the \cite{nettel12b} adiabat at around 100\,MPa.  The lower temperatures require a lower $Z$ in order to fit the density.  This shows that deviations of 20\% or more in the temperature in this region can have noticeable effects on the inferred composition.  The $N=10^6$ (hot) model also requires significantly different values of $Z$, and, indeed, $Z$ oscillates in an unphysical manner.  The reason for this is interesting.  Near the outer surface the small value of $\ell$ means that the temperature must rise quickly due to the inefficiency of convection and the low value of the conductivity.  As a result $Z$ rises as well.  However, when the temperature reaches high enough values, the conductivity of the material increases, and the heat flux can be transported more efficiently.  As a result the temperature gradient decreases and the temperature rises more slowly.  This leads to an eventual decrease in $Z$.  The resulting thermal profile, although generated by a physically plausible argument, does not provide a self-consistent interpretation of the envelope density profile, but it illustrates the importance of a careful calculation of the heat transport.  The composition of the shell and core of Neptune are much less sensitive to the thermal profile.  This is because at the pressures involved, the thermal perturbation to the pressure is small in this region.  It is also interesting to note that because both $Z$ and $T$ are high, the conductivity is high as well, and the innermost region is conductive for both the $N=1$ and $N=10^6$ profiles.  

For model N1 of \cite{nettel12b} the density and temperature profiles were derived simultaneously, so it is not overly surprising that deviations from that temperature profile lead to inconsistencies in the interpretation of the inferred composition.  The more interesting question is whether it is possible to generate an alternative density profile that fits all the measured parameters of the gravity field and yields a reasonable composition with a physically plausible thermal profile.  As a simple case we consider the polynomial profile of \cite{helledetal10}.  We used our thermal transport code to compute a ``cold" profile with $N=1$ and a ``hot" profile with $N=10^6$, and took $r_c=R$.  For this case too we took $R_{Z1}$ and $R_{Z2}$ to be $7.5\times 10^8$ and $2.1\times 10^9$\,cm respectively.  We also ran models with $N=10^6$ and with $R_{Z1}$, $R_{Z2}$, and $r_c$ varying by different amounts, and although the resulting thermal profiles were different, the estimated values of $Z$ did not differ in any profound way.  Table\,\ref{tab:polynep} summarizes the results for the cold and hot profiles for the polynomial density profile. In both cases the results are for the SESAME EOS with \h2o as the high-Z material.  

As can be seen from Table\,\ref{tab:polynep}, the cold model implies that the value of $Z$ oscillates in the planetary interior.  Such a behavior is hard to reconcile on physical grounds, and shows that the polynomial density distribution is not consistent with this temperature profile.  The hot model also displays oscillations in $Z$ as a function of $r$, but they are much smaller.  It is likely that adjustments in the choice of $N$, $R_{Z1}$, $R_{Z2}$, and $r_c$ could produce a thermal profile that would give a physically consistent composition which would provide a very different internal structure for Neptune.  This model is convecting down to a pressure of 180\,GPa ($T=7100$\,K) and then remains diffusive until the center.

\section{Uranus}
\begin{figure}[ht]
\centerline{\includegraphics[angle=0, width=17cm]{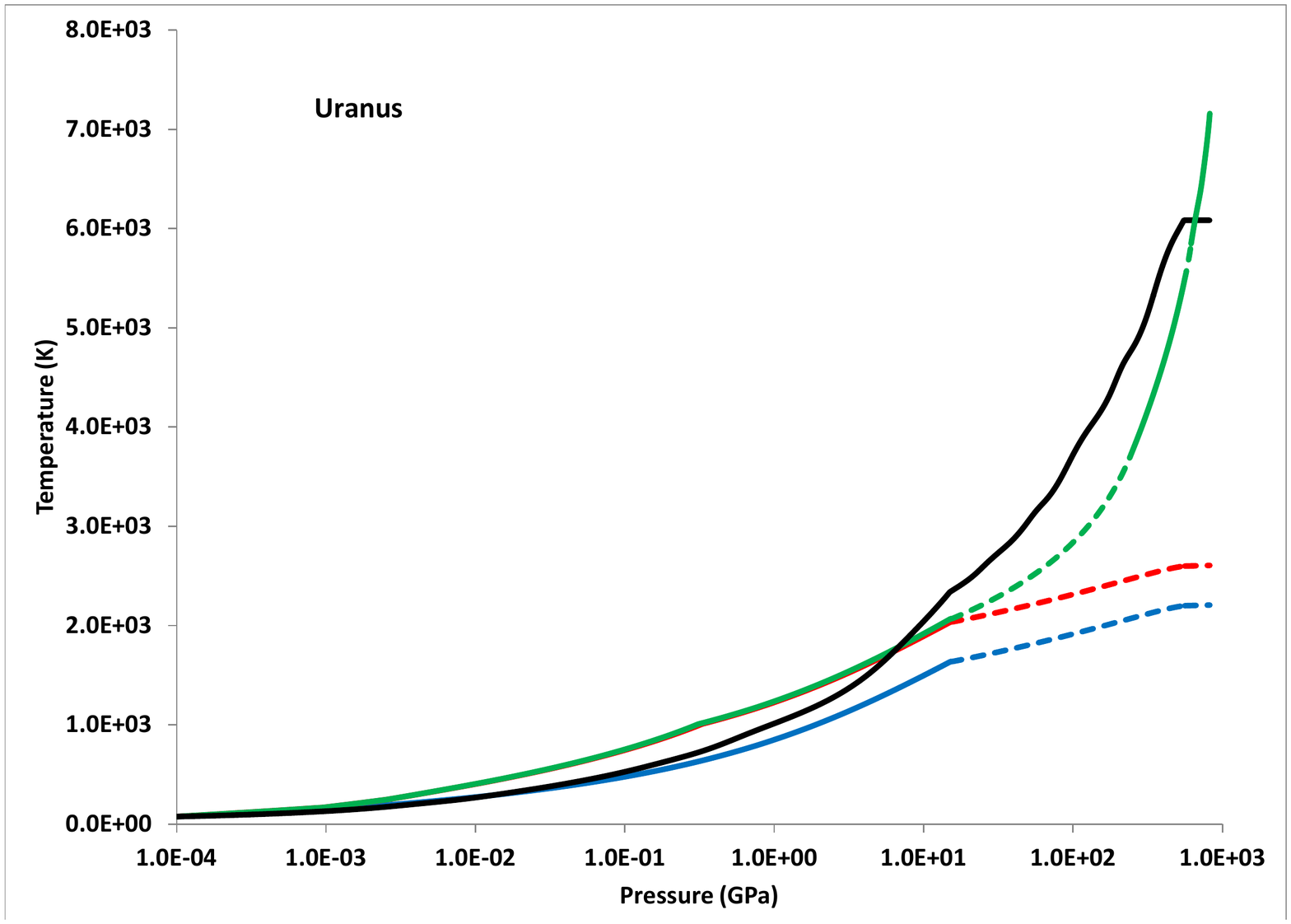}}
\caption{Thermal profiles as a function of internal pressure for Uranus models.  Shown are a 1 layer (blue) and a $10^6$ layer (red) model with $r_c=R$ and a $10^6$ layer model with $r_c=0.1R$ (green).  The temperature profile found by \cite{nettel12b} is shown in black.  The solid parts of the curves are convective regions and the dashed parts are conductive. See text for details. }\label{turn}
\end{figure}

Uranus is a particularly interesting case for our study due to its low thermal flux.  As a result, we can anticipate that large portions of the interior are diffusive, and the thermal profile could be very different from the adiabatic one.  As before, we begin with the model (U1) of \cite{nettel12b} and compute a cold model ($N=1$) and a hot model ($N=10^6$).  However, because Uranus' thermal flux is so low, most of the planet is diffusive and the thermal profile is much less sensitive to the choice of $N$.  Indeed, as can be seen from Fig.\,\ref{turn} the thermal profiles are quite similar and significantly lower than the adiabatic profile derived by \cite{nettel12b}.  

Uranus can be made considerably hotter if we assume that there is some mechanism for trapping the heat at radii less than $r_c$.  Thus, if we set $r_c=0.1R$ we get significantly higher temperatures.  In addition, the thermal transport is significantly different.  For the models with $r_c=R$ only the region with $P<15$\,GPa is convective.  For the model with $r_c=0.1R$ the same outer region is convective, but a second convective layer forms deeper in the planet, at $P=240$\,GPa and continues down to $P=560$\,GPa.  At 620\,GPa a third convective region begins which continues to the center of the planet.  This structure also yields much higher central temperatures.  These thermal profiles are shown in Fig.\,\ref{turn}.  Table\,\ref{tab:nadu} summarizes our results for this density distribution assuming the SESAME EOS.  

As can be seen from Table\,\ref{tab:nadu}, the \citep{nettel12b} model gives a self-consistent picture.  The envelope requires $Z\sim 0.2$.  Because of Uranus' low thermal flux, the thermal profiles computed in our models are considerably colder, and much of the planet is not convective.  This provides an interesting case study for the sensitivity of inferred composition to the thermal structure.  We note that the outer envelope is particularly sensitive to the assumed temperature, as expected.  The inferred value of $Z$ can differ by more than a factor of 2 for the different profiles.  However, the hot profile requires $Z\sim 0.4$ near the outer edge of the envelope and $Z\sim 0.2$ near the shell-envelope boundary, which is probably unrealistic.  As before, the inferred values of $Z$ for the shell and the core are much less sensitive to the temperature. 

\begin{figure}[ht]
\centerline{\includegraphics[angle=0, width=17cm]{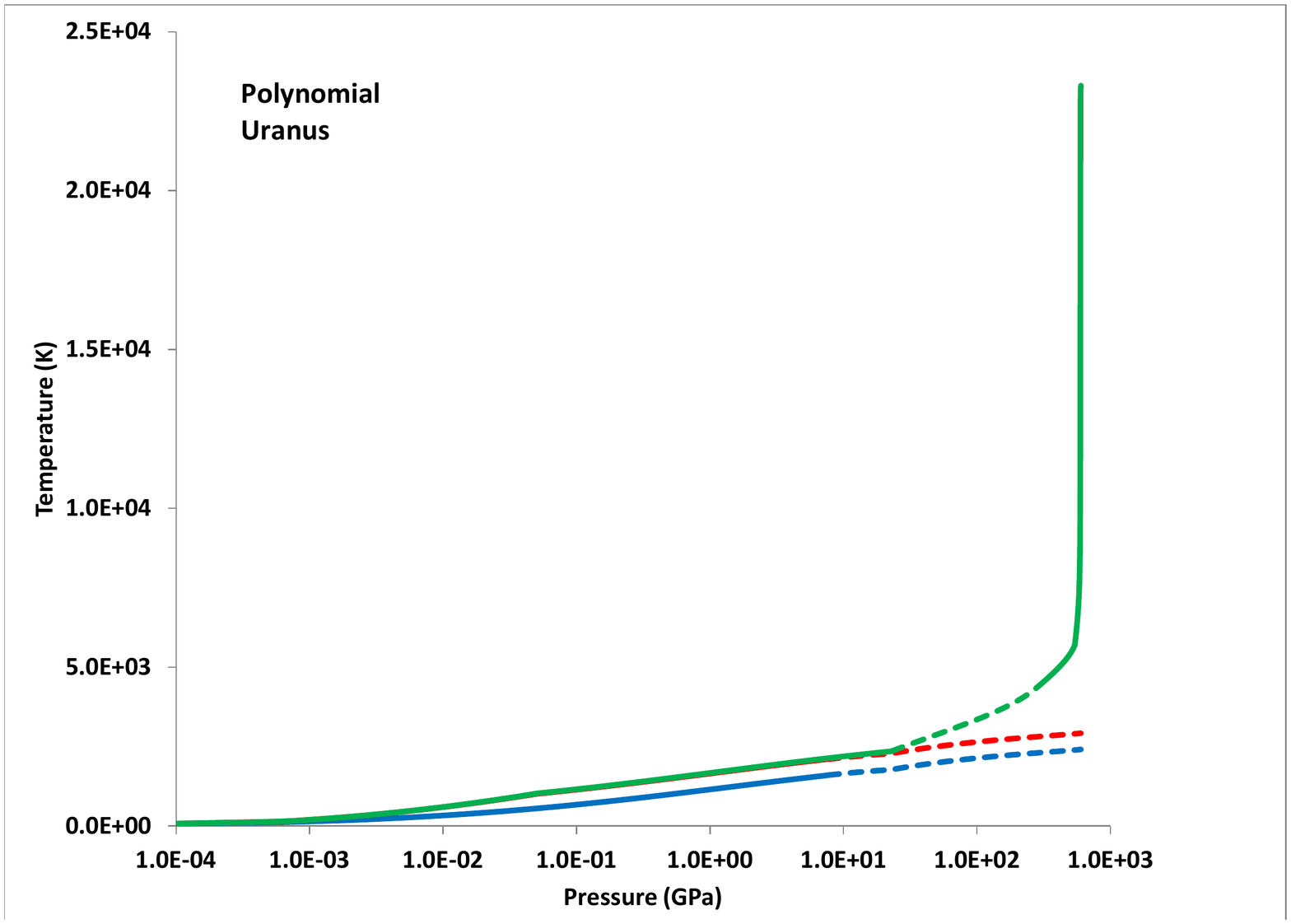}}
\caption{Thermal profiles as a function of internal pressure for the polynomial Uranus models of \cite{helledetal10}.  Shown are a 1 layer (blue) and a $10^6$ layer (red) model with $r_c=R$ and a $10^6$ layer model with $r_c=0.1R$ (green).  The solid parts of the curves indicate convective regions and the dashed parts are conductive.  See text for details. }\label{purn}
\end{figure}

Next we investigated the polynomial model of \cite{heletal10} for comparison.  We took the same values of $R_{Z1}$ and $R_{Z2}$ as for the \cite{nettel12b} models, and $r_c=R$.  We ran a case for $N=1$ (cold), and one for $N=10^6$ (warm).  Because of the low thermal flux, the thermal profiles are relatively similar, with a central temperature of $2.41\times 10^3$\,K in the first case, and $2.92\times 10^3$\,K in the second.  In order to generate a hotter profile, we ran a case with $N=10^6$ and $r_c=0.01R$ (hot).  This produced a profile very similar to the warm profile for pressures under around 30\,GPa, with a much steeper rise for higher pressures, reaching a central temperature of $2.33\times 10^4$\,K. The temperature profiles are shown in Fig.\,\ref{purn}.  The inferred value of $Z$ assuming that the high-Z material is \h2o, is shown in Table\,\ref{tab:polyu} for the cold and hot cases.  

We find that the cold model is not physical as  $Z$ increases from 0.7 to nearly 1 and then becomes negative before gradually increasing again.  The hot model, on the other hand, gives a much more reasonable composition.  Admittedly, the value of 0.7 at 1 bar is high, but this is an artifact of the polynomial approximation which gives a somewhat higher density in the outermost regions of the planet.  A reduction of the 1-bar density by a factor of 2, and the 3 bar density by 20\%, which would have a negligible effect on the gravitational parameters, brings the value of $Z$ down to $\sim 0.35$ throughout the entire region with pressures below $\sim 1$\,GPa.  After that, $Z$ rises gradually to a value of 0.96 at the center.  This central value is not overly sensitive to temperature.  For the cold model, with a central temperature a factor of 10 lower, $Z$ only decreases to 0.86.  The hot model with a polynomial density distribution is certainly a physically plausible alternative to the usual models for Uranus' interior.

\section{Summary and Discussion}
Interior models of the giant planets in our solar system are used to derive the planetary composition by fitting to the observed mass, radius, and gravitational moments using detailed equations of state. In order to do this,  the temperature profile must be known, and typically an adiabatic temperature profile is assumed.  However, for Uranus and probably Neptune, the observed heat flux can be carried through large volumes of the planet by conduction.  We have used published density profiles for Uranus and Neptune combined with the formalism of LC12 to compute plausible non-adiabatic thermal profiles for these bodies. To calculate the temperature profiles, we considered two extreme models: a 1-layer model that gives an approximately adiabatic profile through the convecting region, and a $10^6$-layer model that allows for strong inhibition of the convective motions.  To investigate even higher central temperatures we also considered a case where the central core is still cooling.  We have then used these different temperature profiles together with an equation of state to investigate the sensitivity of the inferred composition of Uranus and Neptune to the temperature profile. For Neptune, with the density distribution of \cite{nettel12b}, we find that the usual adiabatic profile gives a physically consistent description of the variation of $Z$ in the interior.  It is likely that by varying the parameters of the LC formalism it will be possible to find a thermal profile that will give a physically consistent interpretation to the polynomial density distribution of \cite{helledetal10} as well.  While the model of \cite{nettel12b} gives an overall value 0f $Z\sim 0.8$, the model of \cite{heletal10} gives $Z\sim 0.9$.  The value of $Z$ in the hydrogen-rich envelope is much more sensitive to the temperature profile, however, and can easily vary by factors of two depending on the details of the thermal variation.

For Uranus, with its low internal heat flux, we find that although the adiabatic thermal profile of \cite{nettel12b} gives a physically consistent variation of $Z$ in the interior, the polynomial model of \cite{helledetal10} does as well, provided that we assume a region near the center where the heat flow is inhibited.  Such a mechanism has recently be suggested by \cite{Nett2016}.  Here too the model of \cite{nettel12b} gives an overall value 0f $Z\sim 0.8$, the model of \cite{heletal10} gives $Z\sim 0.9$.  For Uranus we find that as a result of the low internal heat flux a large fraction of the volume should be conducting, with the convective region being mostly in the outer layers. This has implications for understanding planetary dynamos and magnetic field generation. Indeed, there are significant differences between the dynamos of the gas and icy planets \citep[e.g.][]{soderlund13}.  Unlike the axially-dipolar magnetic fields of Jupiter and Saturn, the magnetic fields of Uranus and Neptune are non-axisymmetric and highly mulitpolar.  The multipolar character suggests that these dynamos are generated in highly electrically conducting regions (ionic) near the surface. The regions are perhaps thin. Indeed the morphology of Uranus' dynamo can be explained by magnetic field generation in a thin-shell of radius of $\sim 0.8R$ with a non-convective inner region \citep{stanblox04,stanblox06}.  Both Uranus and Neptune must have such regions. However, the deeper parts of the planets could be quite different with Uranus (and maybe Neptune) mostly thermally conductive and stable against convection \citep[see][and references therein]{schubert2011}.   Our inferred thermal structures for Uranus and Neptune are therefore consistent with the current view of dynamo models in these planets.

Our work represents a feasibility study that should be developed further with the next step being a thermal evolution calculation that models the thermal flux more accurately. In addition, additional density-radius relations that are consistent with the observed mass and gravitational moments should be considered. Nevertheless, our study demonstrates the sensitivity of the planetary composition to the assumed temperature profile. In addition, we show that if we forego the assumption of an adiabatic profile, it is possible to find very different density distributions that fit the observed gravity field for Uranus and Neptune. For exoplanetary characterization, this means that estimates of the structure and composition of exoplanets derived under the assumption that the thermal profile is an adiabat represents only a fraction of the possible solutions. We therefore suggest that future studies should consider additional temperature profiles.

\section*{Acknowledgments}
MP and RH would like to acknowledge the hospitality of the International Space Science Institute in Bern, Switzerland.  RH acknowledges support from Swiss National Science Foundation grant 200021 169054.  We all thank the anonymous referee for a careful reading of the manuscript and thoughtful comments.

\bibliographystyle{icarus2}
\bibliography{paper}

\begin{table}
\begin{center}
\begin{tabular}{c|c|c|c|c}
\hline
& {U1}&PolyU& N1 &PolyN \\
\hline
$P$(GPa)& $K$(erg/cm/s/K) & $K$(erg/cm/s/K) & $K$(erg\,cm$^{-1}$\,s$^{-1}$\,K$^{-1}$) & $K$(erg\,cm$^{-1}$\,s$^{-1}$\,K$^{-1}$) \\
\hline
$10^{-3}$ & $1.0\times 10^4$ & $1.0\times 10^4$ & $1.0\times 10^4$ & $1.0\times 10^4$ \\
$10^{-2}$ & $6.4\times 10^5$ & $4.1\times 10^5$ & $1.0\times 10^4$ & $4.2\times 10^5$ \\
$5\times 10^{-2}$ & $1.6\times 10^6$ & $1.1\times 10^6$ & $1.7\times 10^6$ & $1.1\times 10^6$ \\
$10^{-1}$ & $1.7\times 10^6$ & $2.2\times 10^6$ & $2.5\times 10^6$ & $1.7\times 10^6$ \\
$5\times 10^{-1}$ & $4.6\times 10^6$ & $4.7\times 10^6$ & $5.4\times 10^6$ & $4.5\times 10^6$ \\
1 & $6.2\times 10^6$ & $7.1\times 10^6$ & $7.2\times 10^6$ & $6.9\times 10^6$ \\
5 & $1.2\times 10^7$ & $1.9\times 10^7$ & $1.4\times 10^7$ & $1.9\times 10^7$ \\
10 & $1.7\times 10^7$ & $3.0\times 10^7$ & $1.8\times 10^7$ & $2.9\times 10^7$ \\
50 & $3.9\times 10^7$ & $3.2\times 10^7$ & $3.3\times 10^7$ & $3.2\times 10^7$ \\
$10^2$ & $5.1\times 10^7$ & $4.8\times 10^7$ & $4.3\times 10^7$ & $4.8\times 10^7$ \\
$5\times 10^2$ & $9.6\times 10^7$ & $1.1\times 10^8$ & $8.3\times 10^7$ & $1.2\times 10^8$ \\
$8\times 10^2$ & $2.7\times 10^8$ & - & $3.0\times 10^8$ & $1.2\times 10^8$ \\
$1.3\times 10^3$ & - & - & $3.6\times 10^8$ & - \\
\end{tabular}
\caption{Thermal conductivity for different models discussed in the text.  U1 and N1 are the Uranus and Neptune models of \cite{nettel12b} and PolyU and PolyN are, respectively, the polynomial models of Uranus and Neptune.  Blank entries in the table indicate that the respective model does not reach these pressures.}
\label{tab:condt}
\end{center}
\end{table}

\begin{table}
\begin{center}
\begin{tabular}{c|c|c|c|c|c|c|c|c}
\multicolumn{4}{c|}{N1 \cite{nettel12b}}&\multicolumn{2}{c|}{Cold Model}&\multicolumn{2}{c|}{Hot Model}& \\
\hline
$P$(GPa)&$\rho^a$(\gcmc)&$T$(K)&$Z$&$T$(K)&$Z$&$T$(K)&$Z$& \\
\hline
$1.0\times 10^{-4}$&$5.62\times 10^{-4}$&72&0.40&72&0.40&72&0.40& \\
$1.0\times 10^{-3}$&$3.37\times 10^{-3}$&122&0.35&132&0.42&213&0.62& \\
$1.1\times 10^{-2}$&$1.62\times 10^{-2}$&254&0.32&248&0.31&918&0.92& \\
$0.10$&$6.17\times 10^{-2}$&496&0.33&391&0.19&$1.23\times 10^3$&0.74&Envelope \\
1.1&0.168&957&0.32&702&0.22&$1.63\times 10^3$&0.50& \\
10&0.394&$1.91\times 10^3$&0.35&$1.25\times 10^3$&0.28&$2.33\times 10^3$&0.39& \\\hline
10&0.874&$1.91\times 10^3$&0.84&$1.25\times 10^3$&0.81&$2.33\times 10^3$&0.87& \\
$1.0\times 10^2$&1.97&$3.67\times 10^3$&0.84&$3.74\times 10^3$&0.84&$4.98\times 10^3$&0.86&Shell \\
$6.0\times 10^2$&3.72&$5.52\times 10^3$&0.78&$6.54\times 10^3$&0.80&$7.83\times 10^3$&0.81& \\
\hline
$6.0\times 10^2$&9.30&$5.52\times 10^3$&0.91&$6.54\times 10^3$&0.92&$7.83\times 10^3$&0.92& \\
$1.0\times 10^3$&11.1&$5.52\times 10^3$&0.90&$6.63\times 10^3$&0.91&$7.92\times 10^3$&0.92&Core \\
$1.6\times 10^3$&13.1&$5.52\times 10^3$&0.90&$6.71\times 10^3$&0.90&$7.99\times 10^3$&0.90& \\
\end{tabular}
\caption{Composition for different thermal profiles of Neptune.  ``Cold Model" refers to a profile derived by assuming the planet consists of $N=1$ convective layer and ``Hot Model" refers to a profile derived assuming the planet is divided into $N=10^6$ convective layers (see text for details).}
\label{tab:nadnep}
\end{center}
\end{table}

\begin{table}
\begin{center}
\begin{tabular}{c|c|c|c|c|c}
\multicolumn{2}{c|}{}&\multicolumn{2}{c|}{Cold Model}&\multicolumn{2}{c}{Hot Model} \\
\hline
$P$(GPa)&$\rho$(\gcmc)&$T$(K)&$Z$&$T$(K)&$Z$ \\
\hline
$1.0\times 10^{-4}$&$4.38\times 10^{-4}$&72&0.23&72&0.23 \\
$1.4\times 10^{-3}$&$2.51\times 10^{-3}$&187&0.23&543&0.82 \\
$1.1\times 10^{-2}$&$8.47\times 10^{-3}$&351&0.69&$1.41\times 10^3$&0.88 \\
$0.11$&$3.50\times 10^{-2}$&664&0.02&$1.98\times 10^3$&0.7 \\
1.0&0.145&$1.10\times 10^3$&0.25&$2.58\times 10^3$&0.61 \\
10&0.587&$1.58\times 10^3$&0.00&$3.17\times 10^3$&0.73 \\
$1.00\times 10^2$&2.15&$4.46\times 10^3$&0.89&$6.27\times 10^3$&0.91 \\
$8.33\times 10^2$&5.15&$7.17\times 10^3$&0.91&$8.95\times 10^3$&0.92 \\
\end{tabular}
\caption{Composition for poynomial density profile for Neptune \citep{helledetal10}.  ``Cold Model" refers to a profile derived by assuming the planet consists of $N=1$ convective layer and ``Hot Model" refers to a profile derived assuming the planet is divided into $N=10^6$ convective layers (see text for details).}
\label{tab:polynep}
\end{center}
\end{table}

\begin{table}
\begin{center}
\begin{tabular}{c|c|c|c|c|c|c|c|c}
\multicolumn{4}{c|}{U1 \cite{nettel12b}}&\multicolumn{2}{c|}{Cold Model}&\multicolumn{2}{c|}{Hot Model}& \\
\hline
$P$(GPa)&$\rho$(\gcmc)&$T$(K)&$Z$&$T$(K)&$Z$&$T$(K)&$Z$&Z-material \\
\hline
$1.0\times 10^{-4}$&$4.49\times 10^{-4}$&76&0.29&76&0.29&76&0.29& \\
$1.1\times 10^{-3}$&$2.81\times 10^{-3}$&136&0.23&156&0.32&179&0.40& \\
$1.0\times 10^{-2}$&$1.20\times 10^{-2}$&269&0.18&269&0.18&398&0.44& \\
$0.11$&$4.95\times 10^{-2}$&537&0.20&481&0.13&759&0.40&Envelope \\
1.0&0.140&$1.02\times 10^3$&0.19&854&0.12&$1.24\times 10^3$&0.26& \\
10&0.344&$2.05\times 10^3$&0.23&$1.50\times 10^3$&0.17&$1.92\times 10^3$&0.22& \\
15&0.405&$2.34\times 10^3$&0.25&$1.64\times 10^3$&0.19&$2.07\times 10^3$&0.23& \\\hline
15&1.19&$2.34\times 10^3$&0.92&$1.64\times 10^3$&0.89&$2.07\times 10^3$&0.91& \\
$1.0\times 10^2$&3.72&$5.52\times 10^3$&0.90&$1.92\times 10^3$&0.88&$2.84\times 10^3$&0.89&Shell \\
$5.5\times 10^2$&4.07&$6.08\times 10^3$&0.87&$2.20\times 10^3$&0.82&$5.46\times 10^3$&0.86& \\ \hline
$5.5\times 10^2$&9.08&$6.08\times 10^3$&0.92&$2.20\times 10^3$&0.90&$5.46\times 10^3$&0.86&Core \\
$8.2\times 10^2$&10.3&$6.08\times 10^3$&0.91&$2.21\times 10^3$&0.90&$7.16\times 10^3$&0.91& \\ 
\end{tabular}
\caption{Composition for different thermal profiles of Uranus.  ``Cold Model" refers to a profile derived by assuming the planet consists of $N=1$ convective layer and ``Hot Model" refers to a profile derived assuming the planet is divided into $N=10^6$ convective layers and taking $r_c=0.1R$ (see text for details).}
\label{tab:nadu}
\end{center}
\end{table}

\begin{table}
\begin{center}
\begin{tabular}{c|c|c|c|c|c}
\multicolumn{2}{c|}{}&\multicolumn{2}{c|}{Cold Model}&\multicolumn{2}{c}{Hot Model} \\
\hline
$P$(GPa)&$\rho$(\gcmc)&$T$(K)&$Z$&$T$(K)&$Z$ \\
\hline
$1.0\times 10^{-4}$&$1.06\times 10^{-3}$&76&0.70&76&0.70 \\
$1.1\times 10^{-3}$&$2.07\times 10^{-3}$&146&0.96&202&0.33 \\
$1.1\times 10^{-2}$&$7.62\times 10^{-3}$&343&-0.11&621&0.40 \\
$0.11$&$3.20\times 10^{-2}$&683&-0.04&$1.17\times 10^3$&0.36\\
1.0&0.136&$1.15\times 10^3$&0.22&$1.67\times 10^3$&0.38 \\
10&0.576&$1.65\times 10^3$&0.61&$2.19\times 10^3$&0.65 \\
$1.00\times 10^2$&2.14&$2.14\times 10^3$&0.86&$3.36\times 10^3$&0.87 \\
$6.00\times 10^2$&4.42&$2.41\times 10^3$&0.86&$2.33\times 10^4$&0.96 \\
\end{tabular}
\caption{Composition for polynomial density profile for Uranus \citep{helledetal10}.  ``Cold Model" refers to a profile derived by assuming the planet consists of $N=1$ convective layer and ``Hot Model" refers to a profile derived assuming the planet is divided into $N=10^6$ convective layers and $r_c=0.01R$ (see text for details).}
\label{tab:polyu}
\end{center}
\end{table}

\end{document}